\documentclass[aps,twocolumn,superscriptaddress,nofootinbib,longbibliography]{revtex4-2}
\usepackage{amsmath,amssymb,amsfonts,amsthm,mathtools,bm}
\usepackage{graphicx}
\usepackage{hyperref}
\usepackage{xcolor}
\usepackage{orcidlink}
\hypersetup{colorlinks=true,linkcolor=blue,citecolor=blue,urlcolor=blue}

\newtheorem{proposition}{Proposition}

\newcommand{\ket}[1]{\left|#1\right\rangle}
\newcommand{\bra}[1]{\left\langle#1\right|}
\newcommand{\Tr}{\operatorname{Tr}}
\newcommand{\Hcal}{\mathcal{H}}
\newcommand{\W}{\mathcal{W}}
\newcommand{\Ecal}{\mathcal{E}}
\newcommand{\Dlock}{\Delta_{\rm lock}}
\newcommand{\modn}{\;{\rm mod}\;n}
\newcommand{\id}{\mathbb{I}}

\begin{document}

\title{Locally Passive, Globally Charged Quantum Batteries:\\
Coherence-Controlled Work and the Robustness of the Stored Charge}

\author{Asad Ali\orcidlink{0000-0001-9243-417X}} \email{asal68826@hbku.edu.qa}
\affiliation{Qatar Center for Quantum Computing, College of Science and Engineering, Hamad Bin Khalifa University, Doha, Qatar}
\author{Saif Al-Kuwari\orcidlink{0000-0002-4402-7710}}
\affiliation{Qatar Center for Quantum Computing, College of Science and Engineering, Hamad Bin Khalifa University, Doha, Qatar}
\author{James~Q.~Quach\orcidlink{0000-0002-3619-2505}}
\affiliation{The University of Adelaide, SA 5005, Australia}
\date{\today}

\begin{abstract}
A solvable charger--battery model is introduced in which quantum coherence 
controls both where a quantum battery's charge is stored and how robustly 
it survives noise. Charging converts the charger's coherence into 
charger--battery entanglement and splits the deposited work between a 
locally extractable part and a correlation-locked part accessible only 
through joint operations; for a qubit, the split obeys an exact 
complementarity, and at maximal coherence, the battery is locally passive 
with the entire charge locked in correlations. Robustness follows local 
accessibility: the stored energy and locally extractable work are 
population-based, immune to pure dephasing, and limited only by relaxation, 
with an energy half-life, whereas the correlation-locked work is fragile to 
both dephasing and relaxation. Dephasing, global and local depolarization, 
and amplitude damping are treated through a single gain--loss competition 
algebra, and the resulting storage lifetimes are made concrete with 
superconducting-transmon parameters.
\end{abstract}
\maketitle

\section{Introduction}\label{sec:intro}
Quantum batteries are microscopic systems that store energy and deliver it on demand, and a central
promise of the field is that genuinely quantum resources---coherence, entanglement, and collective
operations---can improve how batteries charge and how much work they return
\cite{Alicki2013,Binder2015,Campaioli2017,Ferraro2018,Andolina2019,Campaioli2018,Campaioli2024}. The
figures of merit are concrete: the extractable work, quantified by the \emph{ergotropy}
\cite{Allahverdyan2004,Pusz1978,Lenard1978}; the charging power; and, no less important for any
practical cell, how robustly the charge survives the noise of a real device. Progress on the first
two has been rapid, but robustness---\emph{where} the energy is stored in the many-body state and how
that placement determines its lifetime---has received far less analytic attention, in part because
tractable, closed-form battery models in which one can follow coherence, correlations, and noise
together are scarce.

Correlations sit at the center of this question. Collective charging can be faster than parallel
charging \cite{Binder2015,Campaioli2017,Ferraro2018}, but correlations also change \emph{what} work is
accessible: when a charger and battery are correlated, an agent restricted to local operations can
extract strictly less than an agent with joint control, and the shortfall is work \emph{bound in
correlations} \cite{Skrzypczyk2014,Goold2016,Andolina2019}. The extreme case, in which a bipartite
state is globally active yet each part is locally passive, has been noted from several angles
\cite{PRX5_2015,PRA104_2021,Shi2022,PRA113_2026}, and is usually read as a limitation---energy one
cannot get at. We take the opposite view and ask what such a battery is \emph{good for}: if the charge
lives in correlations rather than in local populations or coherences, how is it charged, and how long
does it last?

This paper answers both questions in a solvable model and arrives at one organizing
statement, exact within this model: \emph{robustness follows local accessibility}. We charge an $n$-level battery from an
$n$-level charger with one entangling stroke, using the charger's quantum coherence as the fuel. The
stroke deposits work that is partitioned---controlled by that coherence---between what a local agent
can extract and what is locked in charger--battery correlations. At maximal coherence the split is
extreme: the reduced battery is left locally passive, so no local operation extracts anything, while
the joint state holds the full charge as \emph{correlation-locked work},
\begin{equation}
\Dlock=\omega\,C_{\ell_1}\qquad(\text{maximal coherence, equally spaced}),
\label{eq:intro_lock}
\end{equation}
released only by a joint operation; under purely local operations it is not extractable output but locally inaccessible---equivalently, protected---work. The input coherence is partitioned between locally extractable and
correlation-locked work through a complementarity that is exact for a qubit
(Sec.~\ref{sec:complementarity}).

What happens next under noise is the central point, and it is dictated by \emph{where} each piece of
the charge is stored. The battery's stored energy and its locally extractable work are functionals of
the energy populations alone; the correlation-locked work draws in addition on the charger--battery
coherences, and only that coherence-enhanced component is exposed to dephasing. Pure dephasing
leaves populations untouched but destroys coherences, so it leaves the stored energy and the local
work \emph{exactly invariant} while degrading the locked work---removing its quantum part. Energy
relaxation, by contrast, drains the populations themselves. The consequence is a separation of
timescales: the battery's energy and locally accessible work are immune to dephasing and decay only
through relaxation, with a lifetime
\begin{equation}
t_{1/2}=T_1\ln 2
\label{eq:intro_halflife}
\end{equation}
set by $T_1$ and independent of the coherence time $T_2$, whereas the correlation-locked work is
fragile to \emph{both} dephasing ($T_\varphi$) and relaxation ($T_1$). Throughout, the charging stroke
is treated as an externally driven unitary whose energetic cost is not included in the ergotropy
budget: the model describes the charged state and the work extractable from it, not a closed
thermodynamic cycle. In this model, the work one can reach locally is the work that survives pure dephasing. This
accessibility--robustness correspondence, illustrated with transmon-scale parameters, is the main conclusion
and the reason the model is worth studying as a battery.

The remaining results support and quantify this picture. To track how realistic noise erodes both the
stored work and the correlations that secure it, we treat four standard channels---dephasing, global
and local depolarization, and amplitude damping---and show that the charger--battery entanglement and
the local ergotropy degrade through a common, compact algebra: each is a competition
$\max[0,\text{gain}-\text{loss}]$, entanglement in the coherences of the state and ergotropy in its
populations. This population-versus-coherence split is why dephasing spares the work while
killing the entanglement; we develop it for arbitrary dimension and
put it to work as a diagnostic for battery degradation rather than as an end in itself. Finally, we map
the four channels onto the measured parameters of superconducting transmons ($T_1$, $T_\varphi$, and
average gate fidelity), making the lifetime \eqref{eq:intro_halflife} and the storage mechanism
concrete for near-term hardware.

The paper is organized as follows. Section~\ref{sec:model} defines the model, the charging
dynamics, and the coherence-to-negativity identity. Section~\ref{sec:work} treats energy, local and
global ergotropy, correlation-locked work, the coherence--work bridge, and the entanglement--work
complementarity.
Section~\ref{sec:unified} is the core: it establishes the unified competition structure, the two
master relations, the corrected sudden-death thresholds, the small-noise fingerprints, the
gain--loss geometry, and the mapping to $T_1$, $T_\varphi$, and gate fidelity with the resulting
self-discharge behavior. Section~\ref{sec:examples} gives qubit and qutrit examples and the reduction to
the two-qubit benchmark. Sections~\ref{sec:discussion} and \ref{sec:conclusion} discuss and
conclude. Appendices collect the partial-transpose spectra, the ergotropy-death derivation, and the
numerical validation.

\section{High-Dimensional Charger--Battery Model}\label{sec:model}
Let $A$ (charger) and $B$ (battery) be $n$-level systems,
$\Hcal_A\cong\Hcal_B\cong\mathbb{C}^n$, with computational basis $\{\ket{0},\ldots,\ket{n-1}\}$ and
$\ket{jk}\equiv\ket{j}_A\otimes\ket{k}_B$. The charger is prepared in the coherent input
\begin{equation}
\ket{\psi_A}=\sum_{j=0}^{n-1}c_j\ket{j}_A,\qquad q_j\equiv|c_j|^2,\quad\sum_jq_j=1,
\label{eq:input}
\end{equation}
and the battery in the ground state $\ket{0}_B$, so $\ket{\Psi_0}_{AB}=\ket{\psi_A}\otimes\ket{0}_B$.

\emph{Hamiltonian.} The battery Hamiltonian is
\begin{equation}
H_B=\sum_{j=0}^{n-1}\epsilon_j\ket{j}\bra{j}_B,\qquad
0=\epsilon_0<\epsilon_1<\cdots<\epsilon_{n-1},
\label{eq:HB}
\end{equation}
and the charger is assigned the same spectrum, $H_A=\sum_j\epsilon_j\ket{j}\bra{j}_A$, so that the
free Hamiltonian is $H_{AB}=H_A+H_B$. The strict ordering avoids degeneracy ambiguities in the
passive state.

\begin{figure}[t]
\centering
\includegraphics[width=\columnwidth]{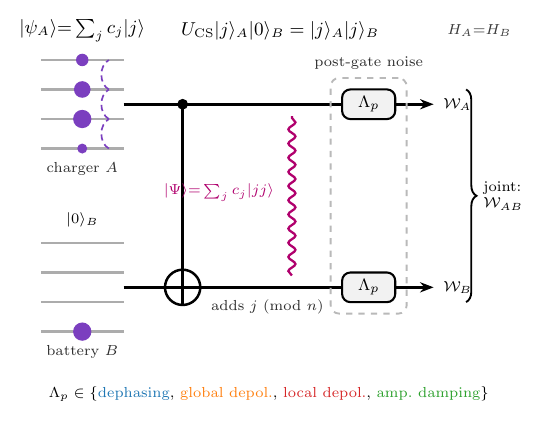}
\caption{\label{fig:model}Model and protocol (drawn for $n=4$). An $n$-level charger $A$, prepared in
$|\psi_A\rangle=\sum_j c_j|j\rangle$, and an $n$-level battery $B$ in its ground state interact through
the controlled modular shift $U_{\rm CS}|j\rangle_A|0\rangle_B=|j\rangle_A|j\rangle_B$. The output
$|\Psi\rangle=\sum_j c_j|jj\rangle$ is maximally correlated: the marginals are diagonal,
$\rho_A=\rho_B=\sum_j q_j|j\rangle\!\langle j|$ (passive at the maximally coherent input,
$\W_A=\W_B=0$), while globally the pair is entangled, $N_0=\tfrac12C_{\ell_1}$, and charged,
$\W_{AB}=2E_B$, with locked work $\Dlock=2E_B^{\rm pas}$. Storage then proceeds under one of four
canonical channels---dephasing, global depolarization, independent local depolarization, and
amplitude damping---whose different actions on the coherence and population sectors determine the
fate of the entanglement and of the extractable work;
the amplitude-damping work threshold quoted is the qubit case.}
\end{figure}
\emph{Charging unitary.} Let $X_n$ be the cyclic shift \emph{on the battery} $\Hcal_B$,
$X_n\ket{k}_B=\ket{k+1\modn}_B$, so its $j$-th power obeys $X_n^j\ket{k}_B=\ket{k+j\modn}_B$; $X_n$
has no action on $\Hcal_A$. The charging stroke is the controlled modular-shift gate
\cite{NielsenChuang2000}
\begin{equation}
U_{\rm CS}^{(n)}=\sum_{j=0}^{n-1}\ket{j}\bra{j}_A\otimes X_n^j ,
\label{eq:UCS}
\end{equation}
which deposits into the battery an amount of energy conditioned on the charger level:
$U_{\rm CS}^{(n)}\ket{j}_A\ket{k}_B=\ket{j}_A\ket{k+j\modn}_B$. Since $U_{\rm CS}^{(n)}\ket{j0}=\ket{jj}$,
\begin{align}
\ket{\Psi}_{AB}&=U_{\rm CS}^{(n)}\ket{\Psi_0}_{AB}=\sum_{j=0}^{n-1}c_j\ket{jj},\nonumber\\
\rho_{AB}&=\sum_{j,k}c_jc_k^*\ket{jj}\bra{kk}.
\label{eq:charged}
\end{align}
The reduced states are identical and diagonal,
$\rho_A=\rho_B=\sum_jq_j\ket{j}\bra{j}$: the populations $q_j$ become local energy populations,
while the phases of $c_j$ become nonlocal coherences in $\rho_{AB}$.

\emph{Reachability and noise protocol.} The charged state \eqref{eq:charged} is reached from the
product state $\ket{\Psi_0}$ by the unitary stroke \eqref{eq:UCS}; it is not postulated. The gate is
treated as an externally driven operation and, since $U_{\rm CS}^{(n)}$ generally does not commute
with $H_{AB}$, its implementation cost lies outside the ergotropy budget---the model describes the
state \emph{after} charging and the work extractable from it, not a complete thermodynamic cycle. Of the results below, the coherence--negativity identity and all battery-local quantities are independent of the charger spectrum, whereas $E_{AB}=2E_B$, $\W_{AB}=2E_B$, and hence $\Dlock=2E_B^{\rm pas}$ rely on the symmetric choice $H_A=H_B$; for $H_A\neq H_B$ they become $E_A+E_B$ and $E_A^{\rm pas}+E_B^{\rm pas}$.
Decoherence during storage or readout is modelled by a completely positive trace-preserving (CPTP)
map $\Ecal$ applied after the stroke,
\begin{equation}
\rho_{\rm noisy}=\Ecal\!\left(U_{\rm CS}^{(n)}\,\rho_0\,U_{\rm CS}^{(n)\dagger}\right),
\qquad \rho_0=\ket{\Psi_0}\bra{\Psi_0},
\label{eq:protocol}
\end{equation}
so that every noisy state analyzed below is explicitly reachable from a product initial state under
charging followed by a physical channel. Figure~\ref{fig:model} summarizes the model and protocol.

 A single strength $p\in[0,1]$ parametrizes each channel
($p=0$ noiseless; $p=1$ fully decohered/relaxed): operationally, $p$ is the probability that the
corresponding elementary error---a phase-randomizing event, a depolarizing event, or a decay---has
acted during the stroke or storage interval, and Sec.~\ref{sec:hardware} maps $p$ onto laboratory
coherence times and gate fidelities.

\emph{Coherence-to-negativity identity.} The input $\ell_1$-coherence in the computational basis is
$C_{\ell_1}=\sum_{j\neq k}|c_jc_k^*|=2\sum_{j<k}|c_jc_k|$
\cite{Baumgratz2014,StreltsovRMP2017}. Entanglement is quantified by the negativity
\cite{Horodecki2009,Peres1996,VidalWerner2002,Plenio2005}. The partial transpose of \eqref{eq:charged} over $B$ is
$\rho_{AB}^{T_B}=\sum_{j,k}c_jc_k^*\ket{jk}\bra{kj}$; the diagonal $\ket{jj}\bra{jj}$ carries
eigenvalue $q_j$, and each pair $\{\ket{jk},\ket{kj}\}$ ($j<k$) forms the block
$\big(\begin{smallmatrix}0&c_jc_k^*\\c_j^*c_k&0\end{smallmatrix}\big)$ with eigenvalues
$\pm|c_jc_k|$. Hence
\begin{equation}
N_0=\sum_{j<k}|c_jc_k|=\tfrac{1}{2}C_{\ell_1}.
\label{eq:N0}
\end{equation}
Equation~\eqref{eq:N0} is an exact identity for the pure, maximally correlated controlled-shift
output; it is the specific instance of the coherence-to-entanglement conversion of
Refs.~\cite{Streltsov2015,Ma2016,Killoran2016,PRA2018,ChinPhysLett2026} for this construction, not a new
resource-theoretic law, and it is generally altered once mixedness precedes the stroke.

\section{Energy, Ergotropy, and Correlation-Locked Work}\label{sec:work}
\subsection{Local ergotropy and passivity}
The battery's stored energy is $E_B=\Tr(\rho_BH_B)=\sum_jq_j\epsilon_j$. The ergotropy of $\rho$ with
Hamiltonian $H$ is $\W(\rho,H)=\Tr(\rho H)-\min_U\Tr(U\rho U^\dagger H)$; the minimizing unitary
places the largest eigenvalue of $\rho$ on the lowest level and so on. With $q_j^\downarrow$ the
populations in nonincreasing order,
\begin{equation}
E_B^{\rm pas}=\sum_jq_j^\downarrow\epsilon_j,\qquad \W_B=E_B-E_B^{\rm pas}.
\label{eq:WB}
\end{equation}
We recall the standard passivity criterion \cite{Pusz1978,Lenard1978,Allahverdyan2004}, used only to
sharpen the contrast with entanglement.
\begin{proposition}[Standard passivity of diagonal states]\label{prop:passive}
For a diagonal state in the strictly increasing energy basis, $\W_B=0$ iff
$q_0\ge q_1\ge\cdots\ge q_{n-1}$.
\end{proposition}
\noindent The point is structural: entanglement \eqref{eq:N0} depends on the pairwise products
$|c_jc_k|$ of amplitudes (off-diagonals), whereas local ergotropy \eqref{eq:WB} depends only on the
ordering of the populations $q_j=|c_j|^2$ against the ladder. The same amplitudes feed two resources
through different functionals, which is why maximal entanglement and zero local ergotropy can
coexist.

\subsection{Global ergotropy and locked work}
A local agent, restricted to independent operations on charger and battery, can extract at most
$\W_A+\W_B$; an agent with joint control extracts the global ergotropy $\W_{AB}$. The
\emph{correlation-locked work} is the difference---the charge that is present but locally
inaccessible,
\begin{equation}
\Dlock=\W_{AB}-\W_A-\W_B\ \ (\ge 0),
\label{eq:Dlock_def}
\end{equation}
distinct from the single-system passive-state ``locked energy'' of Ref.~\cite{PRB2024} in being a
bipartite, global-minus-local quantity. The joint state \eqref{eq:charged} is pure, so its global
passive state places its only nonzero
eigenvalue on the ground level $\ket{00}$ and
$\W_{AB}=E_{AB}=\bra{\Psi}H_{AB}\ket{\Psi}=2\sum_jq_j\epsilon_j=2E_B$. Since $\rho_A=\rho_B$ implies
$\W_A=\W_B$,
\begin{equation}
\Dlock=2E_B-2(E_B-E_B^{\rm pas})=2E_B^{\rm pas}.
\label{eq:Dlock}
\end{equation}
Locked work equals twice the local passive-state energy: it is large precisely when the local states
are energetic but passive. The numerical coincidence with $E_B^{\rm pas}$ is a feature of the
symmetric pure-state structure and should not be conflated with the concept of
Ref.~\cite{PRB2024}: $\Dlock$ is a bipartite global-minus-local gap.

\subsection{Maximally coherent charging and the coherence--work bridge}
For the uniform input $\ket{\psi_A^{\rm MC}}=n^{-1/2}\sum_je^{i\phi_j}\ket{j}$ one has $q_j=1/n$ and
$|c_jc_k|=1/n$, so
\begin{equation}
N_0^{\max}=\frac{n-1}{2},\quad \rho_B^{\rm mm}=\frac{\id_n}{n},\quad \W_A=\W_B=0.
\label{eq:MC}
\end{equation}
Here ``MC'' labels the maximally coherent \emph{input} and ``mm'' the resulting maximally mixed
\emph{battery}: the charger coherence becomes nonlocal entanglement, and the reduced battery has no
coherence in its own basis. The maximally mixed battery is passive under any strictly ordered
Hamiltonian, so $\W_A=\W_B=0$ while $\W_{AB}=2E_B^{\max}>0$: the work is entirely correlation-locked.
The global quantities are $E_B^{\max}=n^{-1}\sum_j\epsilon_j$ and
$\W_{AB}^{\max}=\Dlock^{\max}=2n^{-1}\sum_j\epsilon_j$; for $\epsilon_j=j\omega$,
$E_B^{\max}=\omega(n-1)/2$ and $\Dlock^{\max}=\omega(n-1)$, whereas fixed bandwidth
$\epsilon_j=j\Omega/(n-1)$ gives $\Dlock^{\max}=\Omega$ independent of $n$. Increasing dimension and
increasing bandwidth are distinct resources.

Because $C_{\ell_1}^{\rm MC}=n-1$ for the maximally coherent input, the locked work and the
entanglement are set by the \emph{same} coherence:
\begin{equation}
\boxed{\;\Dlock^{\max}=\omega\,C_{\ell_1}=2\omega\,N_0\;}
\qquad(\text{equally spaced, MC}).
\label{eq:bridge}
\end{equation}
Equation~\eqref{eq:bridge} is a bridge between the informational and thermodynamic sectors:
coherence converts to entanglement through $N_0=C_{\ell_1}/2$ and to locked work through
$\Dlock^{\max}=\omega C_{\ell_1}$, with $\Dlock^{\max}=2\omega N_0$. Maximizing input coherence therefore maximizes the
globally accessible (locked) work while minimizing the locally accessible work.

\subsection{Entanglement--work complementarity}\label{sec:complementarity}
The bridge \eqref{eq:bridge} holds at the maximal-coherence point. Away from it, the same coherence is
partitioned between the two sectors according to an exact \emph{complementarity}. For the qubit
charger--battery ($q_1=q$, $H_B=\omega\ket{1}\bra{1}$), $C_{\ell_1}=2\sqrt{q(1-q)}$ and
$\W_B=\omega\max(0,2q-1)$, so
\begin{equation}
C_{\ell_1}^2+\Big(\frac{\W_B}{\omega}\Big)^2\le 1,
\label{eq:complementarity}
\end{equation}
with \emph{equality for every active battery} $q\ge\tfrac12$ (for $q<\tfrac12$ the battery is passive,
$\W_B=0$, and the left side is $C_{\ell_1}^2\le1$, saturated only at $q=\tfrac12$). The charging stroke thus partitions a fixed coherence budget between
nonlocal entanglement and local extractable
work along the unit circle: a maximally entangled pair ($C_{\ell_1}=1$) stores \emph{zero} locally
extractable work, a fully inverted battery ($\W_B=\omega$) carries \emph{zero} entanglement, and every
intermediate charging trades one for the other. In this form the locked work $\Dlock=2E_B^{\rm pas}$ is
the work displaced from the local sector into correlations as coherence is increased. Its operational value presupposes joint charger--battery control; absent such control, $\Dlock$ quantifies work locked away from---and thereby shielded from---any local agent. In arbitrary
dimension the two extremes persist---maximal coherence gives maximal entanglement and a passive
battery, while a population eigenstate gives an active battery and no entanglement---so a trade-off
remains; the simple circular law \eqref{eq:complementarity} is special to $n=2$, and a tight
dimension- and spectrum-dependent bound is left open (numerically,
$(C_{\ell_1}/(n-1))^2+(\W_B/\epsilon_{n-1})^2\le1$ holds for the equally spaced ladder).

\section{Degradation and Robustness of the Stored Charge}\label{sec:unified}
A real battery is charged, then stored, then discharged in the presence of noise. We now ask how the
correlation-stored charge of Sec.~\ref{sec:work} survives four standard error processes---dephasing,
global and local depolarization, and amplitude damping, applied via \eqref{eq:protocol}. Two things
must be tracked: the extractable work itself, and the charger--battery correlations that hold it
(measured by the negativity). Both turn out to be governed by one compact algebra, which we use as a
bookkeeping tool for degradation; the payoff, reached in Sec.~\ref{sec:hardware}, is a clean split in
time---the battery's energy and locally accessible work are immune to dephasing and limited only by
$T_1$, while the correlations and the locked work they secure decay under dephasing as well. We treat
the correlations first (they are the more fragile of the two), then the work, then combine them.

\subsection{Decay of the charger--battery correlations}\label{sec:Nmaster}
\emph{Dephasing.} Independent phase damping sends $\ket{j}\bra{k}\mapsto\sqrt{1-p}\,\ket{j}\bra{k}$
on each subsystem, so the $\ket{jj}\bra{kk}$ coherence is suppressed by $(1-p)$ and populations are
untouched. Each partial-transpose pair block has eigenvalues $\pm(1-p)|c_jc_k|$, giving
$N_{\rm ph}=(1-p)N_0$.

\emph{Global depolarization.} With
$\rho_{\rm dep}=(1-p)\rho_{AB}+\tfrac{p}{n^2}\id_{n^2}$, each pair block becomes
$\big(\begin{smallmatrix}p/n^2&(1-p)c_jc_k^*\\(1-p)c_j^*c_k&p/n^2\end{smallmatrix}\big)$, so
$N_{\rm dep}=\sum_{j<k}\max\{0,(1-p)|c_jc_k|-p/n^2\}$, with per-pair threshold
$p^{(jk)}_{c,N}=n^2|c_jc_k|/(1+n^2|c_jc_k|)$.

\emph{Local depolarization.} The independent single-qudit channel
$\Ecal_p(\sigma)=(1-p)\sigma+\tfrac{p}{n}\id_n$ is applied to each subsystem,
$\rho_{\rm loc}=(\Ecal_p\otimes\Ecal_p)(\rho_{AB})$. Each subsystem suppresses its off-diagonal by
$(1-p)$, so the charger--battery coherence $\ket{jj}\bra{kk}$ is suppressed \emph{quadratically}, by
$(1-p)^2$, while the local admixtures populate the pair diagonal. The $\{\ket{jk},\ket{kj}\}$ block is
$\big(\begin{smallmatrix}\beta_{jk}&(1-p)^2c_jc_k^*\\(1-p)^2c_j^*c_k&\beta_{jk}\end{smallmatrix}\big)$
with $\beta_{jk}=\tfrac{p(1-p)}{n}(q_j+q_k)+\tfrac{p^2}{n^2}$, giving
\begin{equation}
N_{\rm loc}=\sum_{j<k}\max\!\Big\{0,\,(1-p)^2|c_jc_k|-\tfrac{p(1-p)}{n}(q_j+q_k)-\tfrac{p^2}{n^2}\Big\}.
\label{eq:Nloc}
\end{equation}
Unlike global depolarization, the shift is \emph{population dependent} through $q_j+q_k$; for a qubit
it reduces to $\beta=p(2-p)/4$. Because $(1-p)^2\le(1-p)$, local depolarization suppresses the
entanglement-carrying coherence more strongly than the global channel at equal $p$.

\emph{Amplitude damping.} The single-system Kraus set is
$K_0=\ket{0}\bra{0}+\sqrt{1-p}\sum_{a\ge1}\ket{a}\bra{a}$, $K_a=\sqrt p\,\ket{0}\bra{a}$, applied to
both subsystems. The partial transpose splits into ground--excited blocks $\{\ket{0a},\ket{a0}\}$
with eigenvalues $p(1-p)|c_a|^2\pm(1-p)|c_0c_a|$ and excited--excited blocks
$\{\ket{ab},\ket{ba}\}$ ($1\le a<b$) with eigenvalues $\pm(1-p)^2|c_ac_b|$, so
\begin{align}
N_{\rm AD}=&(1-p)\!\sum_{a\ge1}\max\{0,|c_0c_a|-p|c_a|^2\}\nonumber\\
&+(1-p)^2\!\!\sum_{1\le a<b}\!\!|c_ac_b|.
\label{eq:NAD}
\end{align}
All four results collapse into a single \emph{negativity master relation},
\begin{equation}
N(x)=\sum_{\text{pairs }P}\max\!\big[0,\;f_P(x)\,\kappa_P-g_P(x)\big],\qquad \kappa_P\equiv|c_jc_k|,
\label{eq:master}
\end{equation}
with the channel- and pair-dependent coherence-suppression factor $f_P$ and spectral-shift term
$g_P$ collected in Table~\ref{tab:master}. Entanglement of a pair dies when its shift overtakes its
suppressed coherence, $g_P=f_P\kappa_P$; the overall negativity vanishes when even the strongest
pair ($\kappa=M\equiv\max_P\kappa_P$) dies---the finite-noise entanglement sudden death of
Refs.~\cite{YuEberly2004,YuEberly2009,Almeida2007}, here resolved pair by pair. Equation~\eqref{eq:master} contains the two-qubit ($n=2$) case as a special instance: there a single pair with
$\kappa=\tfrac12\sin2\theta$ gives $N=\max[0,f\tfrac12\sin2\theta-g]$; for general $n$ the qudit charger
contributes $n(n-1)/2$ pairs, and amplitude damping splits them into two structurally different
families.

\begin{table*}[t]
\caption{\label{tab:master}Negativity master relation \eqref{eq:master}: coherence-suppression
factor $f_P$ and spectral-shift term $g_P$ for each channel and pair type. Dephasing has $g=0$ (no
finite death); global depolarization adds a constant isotropic shift; local depolarization suppresses
coherence \emph{quadratically} and adds a population-dependent shift (strictly more damaging to
entanglement than the global channel at equal $p$); amplitude damping splits pairs into a
ground--excited family (finite death when $|c_0|<p|c_a|$) and an excited--excited family (death only
at $p=1$).}
\begin{ruledtabular}
\begin{tabular}{lccc}
Channel & Pair type & $f_P$ & $g_P$ \\
\colrule
Dephasing            & all $(j,k)$        & $1-p$     & $0$ \\
Global depol.        & all $(j,k)$        & $1-p$     & $p/n^2$ \\
Local depol.         & all $(j,k)$        & $(1-p)^2$ & $\tfrac{p(1-p)}{n}(q_j{+}q_k)+\tfrac{p^2}{n^2}$ \\
Amplitude damping    & ground--excited $(0,a)$ & $1-p$     & $p(1-p)|c_a|^2$ \\
Amplitude damping    & excited--excited $(a,b)$ & $(1-p)^2$ & $0$ \\
\end{tabular}
\end{ruledtabular}
\end{table*}

\subsection{Decay of the extractable work}\label{sec:Wmaster}
The same channels act on the population sector as follows.

\emph{Dephasing} leaves populations unchanged: $E_B^{\rm ph}=E_B$, $\W_B^{\rm ph}=\W_B$. Energy and
local ergotropy are dephasing invariant.

\emph{Global depolarization} gives $q_j^{\rm dep}=(1-p)q_j+p/n$, so
$q_j^{\rm dep}-q_k^{\rm dep}=(1-p)(q_j-q_k)$: all population differences rescale by $(1-p)$ with
signs preserved. The passifying permutation is unchanged and
\begin{equation}
\W_B^{\rm dep}=(1-p)\W_B .
\label{eq:Wdep}
\end{equation}
Local ergotropy is suppressed multiplicatively with \emph{no finite death}---an active ordering stays
active for all $p<1$---in contrast to the finite entanglement threshold $p^{(jk)}_{c,N}$.

\emph{Local depolarization} produces the \emph{same} reduced battery as the global channel, because
$\Ecal_p^A$ is trace preserving and hence
$\rho_B^{\rm loc}=\Tr_A[(\Ecal_p\otimes\Ecal_p)\rho_{AB}]=\Ecal_p(\rho_B)=(1-p)\rho_B+\tfrac{p}{n}\id_n$,
i.e. $q_j^{\rm loc}=(1-p)q_j+p/n=q_j^{\rm dep}$. Therefore
\begin{equation}
\W_B^{\rm loc}=(1-p)\W_B=\W_B^{\rm dep},
\label{eq:Wloc}
\end{equation}
again with no finite ergotropy death. This is a sharp illustration of the sector-action principle of
Sec.~\ref{sec:principle}: \emph{global and local depolarization act identically on the population
sector} (identical local ergotropy) \emph{but differently on the coherence sector} (suppression
$(1-p)$ versus $(1-p)^2$ and a larger, population-dependent shift), so they yield the same extractable
local work while local depolarization destroys charger--battery entanglement at a strictly lower
noise threshold. Entanglement and ergotropy thus need not track each other even when two channels
share the same energetics.

\emph{Amplitude damping} moves population toward the ground state,
$q_0^{\rm AD}=q_0+p\sum_{a\ge1}q_a$, $q_a^{\rm AD}=(1-p)q_a$, so $E_B^{\rm AD}=(1-p)E_B$ (a genuine
discharge). Because the ordering can reverse, ergotropy can die at finite $p$. This threshold is
\emph{dimension dependent}, a point that requires care. For a qubit with excited population
$q\equiv q_1>\tfrac12$,
\begin{equation}
\W_B^{\rm AD}=\omega\max\{0,\,2(1-p)q-1\},\qquad p^{\rm AD}_{c,\W}=1-\frac{1}{2q},
\label{eq:AD_qubit}
\end{equation}
which reduces to $p_{c,\W}=1/2$ only for the fully inverted input $q=1$. For $n>2$ the extremal
levels crossing at $p=1/2$ does \emph{not} passify the state, because intermediate levels remain
underpopulated relative to the top level; the fully inverted $n$-level input decays as a piecewise
linear $\W_B^{\rm AD}(p)$ with a slope change at $p=1/2$ and vanishes only at $p=1$ (Appendix
\ref{app:erg}). The general death condition is that $q_j^{\rm AD}(p)$ become nonincreasing in $j$;
the qubit value $1/2$ is the $n=2$ special case, correcting a claim that does not hold in higher
dimension.

\emph{Ergotropy retention fraction.} It is convenient to package these results in one dimensionless
figure of merit, the post-charging \emph{ergotropy retention fraction}
\begin{equation}
\eta_{\W}(p)\equiv\frac{\W_B(p)}{\W_B(0)}\,,\qquad \W_B(0)>0,
\label{eq:retention}
\end{equation}
the fraction of the freshly charged, locally extractable work that survives noise of strength $p$; for
$n=2$ the charging stroke is the CNOT gate, so $\eta_\W$ is the post-CNOT-gate sustained-ergotropy
fraction. The four channels give $\eta_\W^{\rm ph}=1$ exactly (all $n$),
$\eta_\W^{\rm dep}=\eta_\W^{\rm loc}=1-p$, and, for the active qubit input,
$\eta_\W^{\rm AD}=\max\{0,\,1-2qp/(2q-1)\}$, vanishing at $p_{c,\W}=1-1/(2q)$; for the fully inverted
qudit, $\eta_\W^{\rm AD}=1-np/(n-1)$ for $p\le\tfrac12$ and $(n-2)(1-p)/(n-1)$ for $p\ge\tfrac12$,
reaching zero only at $p=1$. Because $\W_B(0)=0$ at the balanced input, $\eta_\W$ is read at an
\emph{active} input and complements the normalized negativity $N(p)/N_0$ read at the balanced input;
its initial slopes $\mathrm d\eta_\W/\mathrm dp|_0$ are exactly the ergotropy fingerprints of
Sec.~\ref{sec:fingerprint}. Under storage for a wait time $t$ it becomes the qubit retention curve
$\eta_\W(t)=\max\{0,\,1-2q(1-e^{-t/T_1})/(2q-1)\}$, dephasing-independent by \eqref{eq:deph_inv}.

\subsection{Why work and correlations decay differently}\label{sec:principle}
The two preceding subsections have a common form, and comparing them explains the property that
matters for the battery: the stored work and the correlations holding it decay through \emph{different}
parts of the state. Write a generic resource as a sum of nonnegative residuals,
\begin{equation}
R(x)=\sum_\mu \max\!\big[0,\;G_\mu(x)-L_\mu(x)\big],
\label{eq:generic}
\end{equation}
a competition between a \emph{gain} $G_\mu$ and a \emph{loss} $L_\mu$ over a set of modes $\mu$. The
negativity \eqref{eq:master} is \emph{exactly} of this form---one residual per partial-transpose pair.
The local ergotropy is a competition of the same type: it is the energy stored in population inversion
minus what thermalization removes, $\W_B=\sum_j\epsilon_j(q_j-q_j^\downarrow)$, and whenever the noisy
populations remain inverted across a single dominant transition (the qubit input, or the
fully-inverted $n$-level input studied here) it collapses to a single residual
$\W_B=\epsilon\,\max[0,\,\text{inversion}-\text{thermalization}]$, as in \eqref{eq:AD_qubit}. In both
sectors, then, sudden death is the vanishing of a $\max[0,\text{gain}-\text{loss}]$ residual; the two
resources differ only in \emph{which} data of the state they read:
\begin{itemize}
\item \emph{Coherence sector} (entanglement): modes are pairs $P$, gain $G_P=f_P|c_jc_k|$
(noise-suppressed pairwise coherence), loss $L_P=g_P$ (partial-transpose diagonal shift from mixing
or relaxation).
\item \emph{Population sector} (ergotropy): modes are inversions of the energy ladder, gain
$G=$ excess population of higher over lower levels, loss $L=$ thermalization that flattens the
distribution.
\end{itemize}
A channel is characterized by its \emph{sector action}---how it acts on off-diagonals (coherence)
versus diagonals (populations). Within this state family and for the channels considered, this action fixes the fate of both resources at once, as summarized
in Table~\ref{tab:sector}.

\begin{table*}[t]
\caption{\label{tab:sector}Sector-action principle. Each channel's action on the coherence sector
(off-diagonals) and the population sector (diagonals) determines the fate of entanglement and
ergotropy. ``Finite death'' means a threshold $p_c<1$; ``no finite death'' means the resource
vanishes only at $p=1$.}
\begin{ruledtabular}
\begin{tabular}{lcccc}
Channel & Coherence & Population & $N$ fate & $\W_B$ fate \\
\colrule
Dephasing         & suppress $(1-p)$ & unchanged & $p{=}1$ only & invariant \\
Global depol.     & suppress $(1-p)$ $+$ shift & rescale $(1-p)$ & finite $p_c$ & $p{=}1$ only \\
Local depol.      & suppress $(1-p)^2$ $+$ shift & rescale $(1-p)$ & finite $p_c'{<}p_c$ & $p{=}1$ only \\
Amplitude damp.   & suppress $+$ shift & relax to $\ket{0}$ & mixed$^\ast$ & finite $p_c$ \\
\end{tabular}
\end{ruledtabular}
\footnotesize $^\ast$Ground--excited pairs can die at finite $p$; excited--excited pairs at $p=1$.
Global and local depolarization have \emph{identical} population action (hence identical $\W_B$) but
different coherence action (hence $p_c'<p_c$ for entanglement).
\end{table*}

The principle is predictive rather than descriptive: given only the sector action of a channel, one
reads off whether entanglement and ergotropy die suddenly or continuously, and why the two need not
coincide. Dephasing is pure coherence-sector loss, so it disentangles without discharging.
Amplitude damping is dominated by population-sector loss, so it discharges (kills ergotropy) while
excited--excited entanglement persists. Global and local depolarization provide a direct test of the
principle: they have \emph{identical} population-sector action---both send $q_j\to(1-p)q_j+p/n$, hence
$\W_B^{\rm loc}=\W_B^{\rm dep}=(1-p)\W_B$---yet different coherence-sector action---suppression $(1-p)$
versus $(1-p)^2$ and a larger, population-dependent shift---so local depolarization extinguishes
charger--battery entanglement at a strictly lower threshold while leaving the extractable local work
unchanged. This is the concrete content of the otherwise generic statement that entanglement and work
are ``model dependent'' \cite{Andolina2019,PRA113_2026}: the dependence is governed by which sector
a channel loads, and two channels that load the population sector identically can still load the
coherence sector very differently.

\subsection{Small-noise fingerprints}\label{sec:fingerprint}
Near $p=0$ each resource decays linearly with a channel-specific slope, giving a diagnostic that
requires only the leading behavior. The two resources are, however, probed at \emph{different}
operating points, because they are maximized by different inputs. At the balanced (maximally coherent)
input, where entanglement is maximal and the battery is passive ($\W_B=0$), the negativity slopes are
$\mathrm dN/\mathrm dp|_0/N_0=-1,\;-\tfrac{n+1}{n},\;-2,\;-2\tfrac{n+1}{n}$ for dephasing, global
depolarization, amplitude damping, and local depolarization, reducing to $-1,-\tfrac32,-2,-3$ for
$n=2$; the ordering $1<\tfrac{n+1}{n}\le\tfrac32<2<2\tfrac{n+1}{n}$ holds in every dimension, so the
channel ranking is dimension-independent even though two of the slopes are not. At an active (population-inverted) input, where the battery carries
ergotropy, the ergotropy slopes are distinct: $0$ (dephasing, invariant); $-1$ in units of $\W_B(0)$
for global \emph{and} local depolarization ($\W_B=(1-p)\W_B(0)$); and $-2q/(2q-1)$ for the qubit
amplitude-damping active input from \eqref{eq:AD_qubit}. The entanglement fingerprint (read at the
balanced input) and the ergotropy fingerprint (read at an inverted input) are therefore
\emph{complementary} probes: together they identify the dominant channel and the loaded sector without
reconstructing the full noise curve. Ergotropy alone cannot separate global from local
depolarization (both have slope $-1$), whereas the negativity fingerprint separates all four
channels---the diagnostic value of using both sectors.

\subsection{Geometry of the two decay channels}\label{sec:geometry}
\begin{figure*}[t]
\centering
\includegraphics[width=\textwidth]{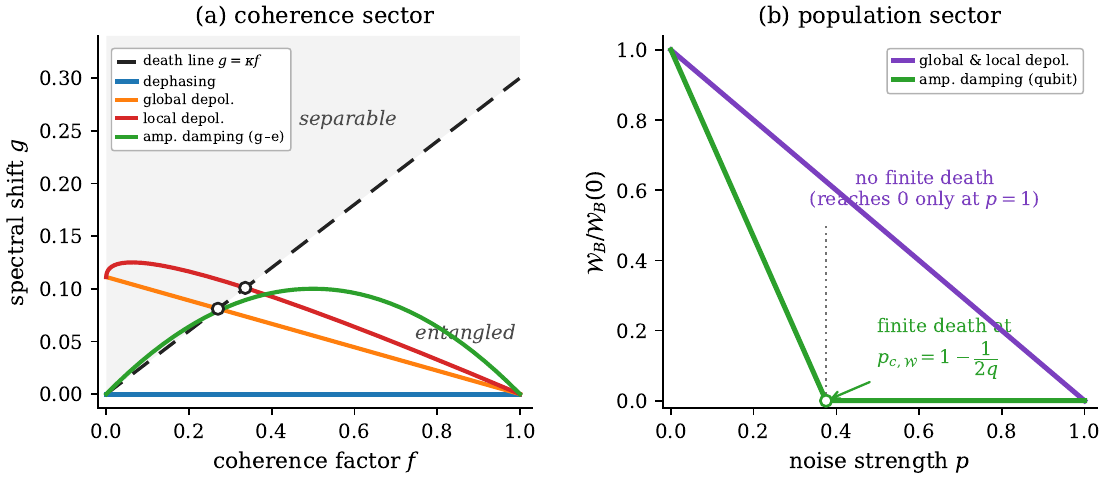}
\caption{\label{fig:geometry}Gain--loss geometry of the unified competition. Left: coherence sector (drawn for $n=3$, $\kappa=0.30$)
in the $(f,g)$ plane; a mode is entangled below the death line $g=f\kappa$ and separable above.
Dephasing (horizontal, $g=0$) never dies at finite $p$; global depolarization moves up the line
$g=(1-f)/n^2$ and crosses the death line at $p^{(jk)}_{c,N}$; local depolarization suppresses
coherence quadratically ($f=(1-p)^2$) with a larger, population-dependent shift and crosses the death
line \emph{earlier}; amplitude damping's ground--excited modes follow the parabola
$g=f(1-f)|c_a|^2$ (vanishing at both endpoints). Right:
population sector (representative qubit case; App.~\ref{app:erg} treats $n>2$); global and local depolarization coincide ($\W_B=(1-p)\W_B(0)$, no finite death),
whereas amplitude damping drives ergotropy to zero at the dimension-dependent threshold
$p_{c,\W}=1-1/(2q)$ (qubit). The two depolarizing channels thus share the population panel but
separate in the coherence panel---the geometric signature of the sector-action principle.}
\end{figure*}

Each coherence mode traces a trajectory in the $(f,g)$ plane as $p$ runs from $0$ to $1$: dephasing
runs along $g=0$ (never entering the separable region); global depolarization moves along the line
$g=(1-f)/n^2$; local depolarization suppresses coherence quadratically, $f=(1-p)^2$, with a larger,
population-dependent shift; amplitude damping's excited--excited modes run along $g=0$ while its
ground--excited modes follow the parabola $g=f(1-f)|c_a|^2$, which vanishes at both $f=1$ ($p=0$) and
$f=0$ ($p=1$). Entanglement of a mode survives while it lies below
the death line $g=f\kappa_P$. The population sector admits the analogous picture in a
(gain,\,loss) plane where the ergotropy-death boundary is the onset of population inversion. Plotting
both boundaries on a common axis (Fig.~\ref{fig:geometry}) exhibits the two sudden-death mechanisms
side by side and makes the sector-action principle geometric: a channel's action in the
coherence plane and in the population plane are independent, so the two deaths are decoupled.

\subsection{Physical realization and self-discharge in superconducting circuits}\label{sec:hardware}
As an order-of-magnitude hardware interpretation, the four channels map onto the standard error processes of superconducting transmon qubits \cite{Krantz2019} and
qutrits, and---because our resources separate into a coherence sector and a population sector---the
mapping predicts qualitatively different storage lifetimes for entanglement and for extractable work.

\emph{Storage errors.} During idle storage the battery experiences energy relaxation and pure
dephasing. Amplitude damping corresponds to $T_1$ relaxation with decay probability
$\gamma=1-e^{-t/T_1}$, and dephasing to the pure-dephasing time $T_\varphi$ with
$p=1-e^{-2t/T_\varphi}$ (the coherence factor is $\sqrt{1-p}=e^{-t/T_\varphi}$), the two combining into
the transverse time $1/T_2=1/(2T_1)+1/T_\varphi$.

\emph{Gate errors.} The charging stroke \eqref{eq:UCS} and the readout are gates whose infidelity is
naturally modelled by depolarization. For a $d$-dimensional depolarizing channel
$(1-p)\rho+p\,\id_d/d$ the average gate fidelity is $\bar F=1-p(d-1)/d$, so the noise strength is fixed
by the measured fidelity: the \emph{local} single-qudit gate ($d=n$) gives
$p_{\rm loc}=\tfrac{n}{n-1}(1-\bar F_1)$ and the \emph{global} two-qudit gate ($d=n^2$) gives
$p_{\rm dep}=\tfrac{n^2}{n^2-1}(1-\bar F_2)$; for qubits these are $p_{\rm loc}=2(1-\bar F_1)$ and
$p_{\rm dep}=\tfrac43(1-\bar F_2)$. We caution that randomized benchmarking reports a related but
differently normalized depolarizing parameter, which should not be interchanged with $p_{\rm dep}$.

\emph{Self-discharge and storage lifetimes.} The consequence is temporal and
depends on where each part of the charge is stored. The battery's stored energy $E_B=\sum_jq_j\epsilon_j$
and its locally extractable work $\W_B=E_B-E_B^{\rm pas}$ are functionals of the populations alone,
whereas the correlation-locked work $\Dlock=\W_{AB}-\W_A-\W_B$ draws on the charger--battery
coherences that enter the global ergotropy $\W_{AB}$. Pure dephasing leaves populations invariant, so
\begin{equation}
E_B(t)=E_B(0),\qquad \W_B(t)=\W_B(0)\quad(\text{dephasing}),
\label{eq:deph_inv}
\end{equation}
the stored energy and the locally accessible work are \emph{exactly immune} to dephasing, even as it
destroys the charger--battery entanglement on the timescale $T_\varphi$ ($N(t)\sim e^{-2t/T_\varphi}N_0$).
The locked work is not immune. For the qubit in the charged regime ($q_1\ge q_0$) it interpolates as
$\Dlock(p)=\omega\,[\,2q_0-\lambda_-(p)\,]$ with
$\lambda_-(p)=\tfrac12-\sqrt{\tfrac14(q_0-q_1)^2+(1-p)^2q_0q_1}$, falling from $2E_B^{\rm pas}$ at
$p=0$ to \emph{exactly half} that value, $E_B^{\rm pas}$, at full dephasing: dephasing removes
precisely the coherence-enhanced half of the locked work and leaves the classically-locked half
carried by the surviving population correlations. Energy relaxation, by contrast, drains the populations
themselves, rescaling the stored energy by $e^{-t/T_1}$,
\begin{equation}
E_B(t)=e^{-t/T_1}E_B(0),\qquad t_{1/2}=T_1\ln 2\quad(\text{relaxation}).
\label{eq:halflife}
\end{equation}
Hence the timescales separate: the battery's energy and locally accessible work are immune to
dephasing and limited only by $T_1$, while the correlation-locked work is fragile to \emph{both}
dephasing ($T_\varphi$) and relaxation ($T_1$). The work one can
reach locally is the work that survives dephasing, and by the complementarity
\eqref{eq:complementarity} coherence trades this robust, accessible work for fragile work stored in
correlations. This is the temporal face of the population-versus-coherence split: energy and local
work live in the populations ($T_1$), correlations and the locked work they secure live in the
coherences ($T_\varphi$).

\emph{Representative numbers (illustrative).} For transmon parameters \cite{Krantz2019} $T_1\approx100\,\mu$s and
$T_\varphi\approx80\,\mu$s one has $T_2\approx57\,\mu$s and an energy half-life
$t_{1/2}\approx69\,\mu$s, while a $200\,$ns stroke incurs only $\gamma\approx2\times10^{-3}$ and
$p\approx5\times10^{-3}$ (perturbative regime); two-qubit gate fidelities $\bar F_2\approx0.99$ give
$p_{\rm dep}\approx1.3\times10^{-2}$. Two caveats specific to qudit hardware: transmon relaxation
cascades down the ladder ($\ket{k}\to\ket{k-1}$) rather than directly to the ground state as in our
amplitude-damping model (the two coincide for $n=2$), and higher levels relax faster
($T_1^{(k)}\!\sim\!T_1/k$), so the qutrit and higher-$n$ estimates are order-of-magnitude reference
points rather than exact device predictions.
For equally spaced levels with cascade rates $\Gamma_k=k\Gamma$, however, $\mathrm d\langle E\rangle/\mathrm dt=-\Gamma\langle E\rangle$ exactly, so the energy law $E_B(t)=e^{-t/T_1}E_B(0)$ and $t_{1/2}=T_1\ln2$ survive the cascade; only the state trajectory, and with it $\W_B(t)$ and its thresholds, becomes device-specific. 

\section{Explicit Examples}\label{sec:examples}
\emph{Qubit ($n=2$).} With $q_0=1-q$, $q_1=q$, $H_B=\omega\ket{1}\bra{1}$,
\begin{equation}
E_B=\omega q,\quad \W_B=\omega\max(0,2q-1),\quad N_0=\sqrt{q(1-q)}.
\end{equation}
At $q=\tfrac12$ the state is maximally entangled but locally passive ($\W_B=0$); for $q>\tfrac12$ the
battery is population inverted and locally active. The negativity master relation reduces exactly, for $n=2$, to
the two-qubit case: with $\kappa=|c_0c_1|=\tfrac12\sin2\theta$ for
$\ket{\psi_A}=\cos\theta\ket0+e^{i\phi}\sin\theta\ket1$, depolarization gives
$N_{\rm dep}=\max\{0,(1-p)\tfrac12\sin2\theta-p/4\}$ and dephasing $N_{\rm ph}=(1-p)\tfrac12\sin2\theta$,
with slope fingerprints $-1,-\tfrac32,-2$. The present model thus
contains this two-qubit case as its $n=2$ instance and extends it to arbitrary dimension and to the
ergotropy sector.

\emph{Qutrit ($n=3$).} With levels $0,\omega,2\omega$,
$E_B=\omega q_1+2\omega q_2$ and $N_0=|c_0c_1|+|c_0c_2|+|c_1c_2|$. If $q_2\ge q_1\ge q_0$ then
$E_B^{\rm pas}=\omega q_1+2\omega q_0$ and $\W_B=2\omega(q_2-q_0)$; other orderings follow from
\eqref{eq:WB}. Even here entanglement and work depend on structurally different combinations: $N_0$
on all three pairwise products, $\W_B$ on the extremal populations $q_0,q_2$ only. Under
\emph{global} depolarization all three pairs share the constant shift $p/9$ and die in order of
increasing $\kappa_P$, while $\W_B$ merely scales by $(1-p)$; under \emph{local} depolarization the
coherence is suppressed by $(1-p)^2$ and each pair carries its own population-dependent shift
$\tfrac{p(1-p)}{3}(q_j+q_k)+\tfrac{p^2}{9}$, so entanglement dies earlier than in the global case even
though $\W_B$ is identical, $\W_B^{\rm loc}=\W_B^{\rm dep}=(1-p)\W_B$; under amplitude damping the
$(1,2)$ pair (excited--excited) outlives the $(0,1),(0,2)$ pairs, and $\W_B$ follows the corrected
dimension-dependent law of Sec.~\ref{sec:Wmaster}.

\section{Discussion}\label{sec:discussion}
\emph{What is new.} The primary result is physical and battery-specific: a solvable model of a
quantum battery in which coherence controls how the charge is stored and how robust it is. The
charging coherence sets the correlation-locked work [reaching $\Dlock=\omega C_{\ell_1}$ at maximal
coherence for an equally spaced battery, Eq.~\eqref{eq:bridge}] and, through a complementarity that is
exact for a qubit \eqref{eq:complementarity}, partitions
work between a locally accessible part and a correlation-locked part. Robustness then tracks
accessibility: the battery's stored energy and its locally accessible work are population-based, hence
immune to pure dephasing and limited only by relaxation [$t_{1/2}=T_1\ln2$, Eq.~\eqref{eq:halflife}],
while the coherence-enhanced component of the correlation-locked work is fragile to dephasing as well as
relaxation. Coherence thus buys global accessibility at the cost of dephasing-robustness. This
accessibility--robustness correspondence, made concrete for transmon parameters, is the paper's
message. The supporting technical contribution is the observation that the two things one must track
during storage---the work and the correlations securing it---degrade through one compact algebra
\eqref{eq:generic}: the individual ingredients (local passivity with global activity
\cite{PRX5_2015,PRA104_2021,Shi2022}, coherence-to-entanglement conversion
\cite{Streltsov2015,Ma2016,PRA2018,ChinPhysLett2026}, and the model dependence of noisy ergotropy
\cite{PRA113_2026,PRL2025,NJP2025}) are known, but the population-versus-coherence split
(Table~\ref{tab:sector}) explains \emph{why} work and entanglement decay differently and organizes
their degradation through one algebra in arbitrary dimension,
here as a diagnostic for battery degradation rather than as an end in itself.

\emph{Operational meaning of locked work.} A local agent maximizes work with population-inverted
inputs, for which increasing coherence is detrimental; a global agent maximizes work with maximally
coherent inputs, which lock the work in correlations. This control-dependent performance map is not a
weakness of coherent charging but a statement of \emph{where} the work resides. Locked work is a
resource whenever joint control is available---collective or cavity-mediated extraction
\cite{Ferraro2018,Andolina2019}---and a security property otherwise: $\Dlock=2E_B^{\rm pas}$ is
the work that a local adversary cannot extract. By Eq.~\eqref{eq:bridge}, coherence sets both the
entanglement and the globally accessible work.
The self-discharge analysis \eqref{eq:halflife} adds the storage constraint: the battery's energy and its locally
accessible work are limited only by $T_1$ (dephasing does not touch them), whereas the
correlation-locked work is limited by both $T_1$ and $T_\varphi$. If the application can use joint
control, coherent charging maximizes deliverable work but exposes it to dephasing; if only local
extraction and long storage are available, a lower-coherence, population-inverted charge sacrifices
locked work for a $T_1$-limited, dephasing-immune charge. Choosing the charging coherence is thus
simultaneously a choice of accessibility and of robustness.

\emph{Positioning.} Relative to Bell-diagonal and locally passive analyses
\cite{PRX5_2015,PRA104_2021,Shi2022}, the present model is a pure, high-dimensional, exactly
solvable realization in which every quantity---negativity, local and global ergotropy, locked work,
and all noise thresholds---is closed form and organized by one algebra. Relative to noisy-ergotropy
studies \cite{PRA113_2026,PRL2025,NJP2025}, it supplies the missing structural reason for the
entanglement--work mismatch: the two resources occupy different sectors that channels load
independently. Relative to coherence-conversion results \cite{PRA2018,ChinPhysLett2026,Ma2016}, it
couples the conversion identity to a thermodynamic partner \eqref{eq:bridge}.

\paragraph*{Limitations.} Our claims are exact within the controlled-shift model, for the four channels considered, and under the stated access assumptions; we do not present them as universal properties of quantum batteries. The charging stroke is externally driven and its energetic cost is not included: the analysis characterizes the post-charging state and its storage robustness, not a complete thermodynamic cycle, a charging efficiency, or a net work balance. Global ergotropy and the locked work presuppose joint charger--battery control. The qudit amplitude-damping channel is the idealized direct-to-ground map; real transmon qudits relax through level-dependent cascades, for which the energy decay law survives when $\Gamma_k=k\Gamma$ but state trajectories become device-specific. The transmon mapping is illustrative---an order-of-magnitude interpretation rather than a device proposal---and we make no claim of universal quantum advantage or of charging-power enhancement.

\section{Conclusion}\label{sec:conclusion}
We introduced a solvable charger--battery model of a quantum battery in which coherence controls both
how the charge is stored and how robust it is. The charging coherence fixes the correlation-locked
work (reaching $\Dlock=\omega C_{\ell_1}$ at maximal coherence for an equally spaced battery) and
splits work between a locally accessible part and a correlation-locked part through the
qubit complementarity \eqref{eq:complementarity}. Robustness
follows accessibility: the battery's stored energy and its locally accessible work are population-based,
hence immune to pure dephasing and limited only by relaxation ($t_{1/2}=T_1\ln2$), while the coherence-enhanced component of the
correlation-locked work is destroyed by dephasing, its classical remainder decaying only by relaxation. To quantify degradation we
tracked both the work and the correlations that secure it through a single algebra: each is a
$\max[0,\text{gain}-\text{loss}]$ competition, entanglement in the coherences and work in the
populations, which explains their different noise responses and applies in arbitrary dimension within this model. Illustrated with transmon
parameters, the model gives concrete, testable predictions for near-term hardware. Natural extensions
include
non-Markovian and correlated noise, restricted (locally constrained) global operations, and a
resource-theoretic optimization of the coherence-to-locked-work conversion.

\section*{Data availability}
No datasets were generated; all analytical results are contained in the manuscript and validated by
the simulations described in Appendix~\ref{app:num}.

\begin{acknowledgments}
The authors acknowledge support from the Qatar Center for Quantum Computing at Hamad Bin Khalifa
University.
\end{acknowledgments}

\appendix
\section{Partial-transpose spectra and master relation}\label{app:PT}
For $\rho_{AB}=\sum_{jk}c_jc_k^*\ket{jj}\bra{kk}$, $\rho_{AB}^{T_B}=\sum_{jk}c_jc_k^*\ket{jk}\bra{kj}$
is block diagonal: diagonal terms $\ket{jj}\bra{jj}$ (eigenvalue $q_j$) and, for $j<k$, the
$2\times2$ block $\big(\begin{smallmatrix}0&c_jc_k^*\\c_j^*c_k&0\end{smallmatrix}\big)$ with
eigenvalues $\pm\kappa_P$. Under dephasing the off-diagonal scales by $(1-p)$; under global
depolarization the block acquires diagonal $p/n^2$; under local depolarization
$(\Ecal_p\otimes\Ecal_p)$ the off-diagonal scales by $(1-p)^2$ and the block acquires diagonal
$\beta_{jk}=\tfrac{p(1-p)}{n}(q_j+q_k)+\tfrac{p^2}{n^2}$ (the two local admixtures $\rho_A\otimes\id$,
$\id\otimes\rho_B$ contribute $\tfrac{p(1-p)}{n}q_j$ and $\tfrac{p(1-p)}{n}q_k$, and the doubly-mixed
term contributes $\tfrac{p^2}{n^2}$); under amplitude damping the ground--excited block acquires
diagonal $p(1-p)|c_a|^2$ and off-diagonal $(1-p)|c_0c_a|$, while the excited--excited block has
off-diagonal $(1-p)^2|c_ac_b|$ and zero diagonal. Collecting the negative eigenvalues gives
Table~\ref{tab:master} and Eqs.~\eqref{eq:master}--\eqref{eq:NAD}.

\section{Amplitude-damping ergotropy death}\label{app:erg}
For the qubit active input $q>\tfrac12$, $q_1^{\rm AD}=(1-p)q$ and $q_0^{\rm AD}=1-(1-p)q$, so
$\W_B^{\rm AD}=\omega\max\{0,2(1-p)q-1\}$ vanishes at $p^{\rm AD}_{c,\W}=1-1/(2q)$. For the fully
inverted $n$-level input ($q_{n-1}=1$), after damping $q_0^{\rm AD}=p$, $q_{n-1}^{\rm AD}=1-p$, and
all intermediate populations remain zero. Sorting the two nonzero populations onto the lowest
energies gives the exact piecewise ergotropy
\begin{equation}
\W_B^{\rm AD}(p)=\epsilon_{n-1}(1-p)-\epsilon_1\min(p,1-p),
\end{equation}
which for an equally spaced spectrum $\epsilon_j=j\omega$ reads
$\W_B^{\rm AD}=\omega[(n-1)-np]$ for $p\le\tfrac12$ and $\W_B^{\rm AD}=(n-2)\omega(1-p)$ for
$p\ge\tfrac12$. It is continuous, changes slope at $p=\tfrac12$, and---because the top level stays
overpopulated relative to the empty intermediate levels---vanishes only at $p=1$ for every $n>2$.
The naive extremal-crossing value $p=\tfrac12$ is therefore the $n=2$ special case; the general
death condition is the onset of a nonincreasing $q_j^{\rm AD}(p)$, confirmed numerically in
Appendix~\ref{app:num}.

\section{Numerical validation}\label{app:num}
All closed-form expressions were checked against direct density-matrix simulation of the protocol
\eqref{eq:protocol} for $n=2$--$6$ and random coherent inputs. The negativity master relation
\eqref{eq:master} (including the local-depolarizing entry \eqref{eq:Nloc}) and the ergotropy laws
\eqref{eq:Wdep}--\eqref{eq:Wloc} and \eqref{eq:AD_qubit} agree with simulation
to machine precision ($\lesssim10^{-15}$) across $p\in[0,1]$ for all four channels.



\begin{thebibliography}{99}
\bibitem{Pusz1978} W. Pusz and S. L. Woronowicz, Passive states and KMS states for general quantum systems, Commun. Math. Phys. \textbf{58}, 273 (1978).
\bibitem{Lenard1978} A. Lenard, Thermodynamical proof of the Gibbs formula for elementary quantum systems, J. Stat. Phys. \textbf{19}, 575 (1978).
\bibitem{Allahverdyan2004} A. E. Allahverdyan, R. Balian, and T. M. Nieuwenhuizen, Maximal work extraction from finite quantum systems, Europhys. Lett. \textbf{67}, 565 (2004).
\bibitem{Alicki2013} R. Alicki and M. Fannes, Entanglement boost for extractable work from ensembles of quantum batteries, Phys. Rev. E \textbf{87}, 042123 (2013).
\bibitem{Binder2015} F. C. Binder, S. Vinjanampathy, K. Modi, and J. Goold, Quantacell: powerful charging of quantum batteries, New J. Phys. \textbf{17}, 075015 (2015).
\bibitem{Campaioli2017} F. Campaioli, F. A. Pollock, F. C. Binder, L. C\'eleri, J. Goold, S. Vinjanampathy, and K. Modi, Enhancing the charging power of quantum batteries, Phys. Rev. Lett. \textbf{118}, 150601 (2017).
\bibitem{Ferraro2018} D. Ferraro, M. Campisi, G. M. Andolina, V. Pellegrini, and M. Polini, High-power collective charging of a solid-state quantum battery, Phys. Rev. Lett. \textbf{120}, 117702 (2018).
\bibitem{Andolina2019} G. M. Andolina, M. Keck, A. Mari, M. Campisi, V. Giovannetti, and M. Polini, Extractable work, the role of correlations, and asymptotic freedom in quantum batteries, Phys. Rev. Lett. \textbf{122}, 047702 (2019).
\bibitem{Campaioli2018} F. Campaioli, F. A. Pollock, and S. Vinjanampathy, Quantum batteries, in \emph{Thermodynamics in the Quantum Regime: Fundamental Aspects and New Directions}, edited by F. Binder, L. A. Correa, C. Gogolin, J. Anders, and G. Adesso, Fundamental Theories of Physics Vol.~195 (Springer, Cham, 2019), pp.~207--225.
\bibitem{Campaioli2024} F. Campaioli, S. Gherardini, J. Q. Quach, M. Polini, and G. M. Andolina, Colloquium: Quantum batteries, Rev. Mod. Phys. \textbf{96}, 031001 (2024).
\bibitem{Skrzypczyk2014} P. Skrzypczyk, A. J. Short, and S. Popescu, Work extraction and thermodynamics for individual quantum systems, Nat. Commun. \textbf{5}, 4185 (2014).
\bibitem{Goold2016} J. Goold, M. Huber, A. Riera, L. del Rio, and P. Skrzypczyk, The role of quantum information in thermodynamics---a topical review, J. Phys. A \textbf{49}, 143001 (2016).
\bibitem{Horodecki2009} R. Horodecki, P. Horodecki, M. Horodecki, and K. Horodecki, Quantum entanglement, Rev. Mod. Phys. \textbf{81}, 865 (2009).
\bibitem{Baumgratz2014} T. Baumgratz, M. Cramer, and M. B. Plenio, Quantifying coherence, Phys. Rev. Lett. \textbf{113}, 140401 (2014).
\bibitem{StreltsovRMP2017} A. Streltsov, G. Adesso, and M. B. Plenio, Colloquium: Quantum coherence as a resource, Rev. Mod. Phys. \textbf{89}, 041003 (2017).
\bibitem{Streltsov2015} A. Streltsov, U. Singh, H. S. Dhar, M. N. Bera, and G. Adesso, Measuring quantum coherence with entanglement, Phys. Rev. Lett. \textbf{115}, 020403 (2015).
\bibitem{Ma2016} J. Ma, B. Yadin, D. Girolami, V. Vedral, and M. Gu, Converting coherence to quantum correlations, Phys. Rev. Lett. \textbf{116}, 160407 (2016).
\bibitem{Killoran2016} N. Killoran, F. E. S. Steinhoff, and M. B. Plenio, Converting nonclassicality into entanglement, Phys. Rev. Lett. \textbf{116}, 080402 (2016).
\bibitem{NielsenChuang2000} M. A. Nielsen and I. L. Chuang, \emph{Quantum Computation and Quantum Information} (Cambridge University Press, Cambridge, 2000).
\bibitem{Peres1996} A. Peres, Separability criterion for density matrices, Phys. Rev. Lett. \textbf{77}, 1413 (1996).
\bibitem{VidalWerner2002} G. Vidal and R. F. Werner, Computable measure of entanglement, Phys. Rev. A \textbf{65}, 032314 (2002).
\bibitem{Plenio2005} M. B. Plenio, Logarithmic negativity: a full entanglement monotone that is not convex, Phys. Rev. Lett. \textbf{95}, 090503 (2005).
\bibitem{YuEberly2004} T. Yu and J. H. Eberly, Finite-time disentanglement via spontaneous emission, Phys. Rev. Lett. \textbf{93}, 140404 (2004).
\bibitem{YuEberly2009} T. Yu and J. H. Eberly, Sudden death of entanglement, Science \textbf{323}, 598 (2009).
\bibitem{Almeida2007} M. P. Almeida, F. de Melo, M. Hor-Meyll, A. Salles, S. P. Walborn, P. H. Souto Ribeiro, and L. Davidovich, Environment-induced sudden death of entanglement, Science \textbf{316}, 579 (2007).
\bibitem{PRX5_2015} M. Perarnau-Llobet, K. V. Hovhannisyan, M. Huber, P. Skrzypczyk, N. Brunner, and A. Ac\'in, Extractable work from correlations, Phys. Rev. X \textbf{5}, 041011 (2015).
\bibitem{PRA104_2021} K. Sen and U. Sen, Local passivity and entanglement in shared quantum batteries, Phys. Rev. A \textbf{104}, L030402 (2021).
\bibitem{PRA113_2026} M. B. Arjmandi, Frozen and growing quantum work under noise: coherence and correlations as key resources, Phys. Rev. A \textbf{113}, 032220 (2026).
\bibitem{Shi2022} H.-L. Shi, S. Ding, Q.-K. Wan, X.-H. Wang, and W.-L. Yang, Entanglement, coherence, and extractable work in quantum batteries, Phys. Rev. Lett. \textbf{129}, 130602 (2022).
\bibitem{PRL2025} R. P. A. Simon, J. Anders, and K. V. Hovhannisyan, Correlations enable lossless ergotropy transport, Phys. Rev. Lett. \textbf{134}, 010408 (2025).
\bibitem{ChinPhysLett2026} C.-J. Wang and F.-Q. Dou, Ergotropy in quantum batteries, Chin. Phys. Lett. \textbf{43}, 060602 (2026).
\bibitem{NJP2025} P.-Y. Sun, H. Zhou, and F.-Q. Dou, Cavity-Heisenberg spin-$j$ chain quantum battery and reinforcement learning optimization, New J. Phys. \textbf{27}, 124513 (2025).
\bibitem{PRB2024} D.-L. Yang, F.-M. Yang, and F.-Q. Dou, Three-level Dicke quantum battery, Phys. Rev. B \textbf{109}, 235432 (2024).
\bibitem{PRA2018} H. Zhu, M. Hayashi, and L. Chen, Axiomatic and operational connections between the $l_1$-norm of coherence and negativity, Phys. Rev. A \textbf{97}, 022342 (2018).
\bibitem{Krantz2019} P. Krantz, M. Kjaergaard, F. Yan, T. P. Orlando, S. Gustavsson, and W. D. Oliver, A quantum engineer's guide to superconducting qubits, Appl. Phys. Rev. \textbf{6}, 021318 (2019).
\end{thebibliography}
\end{document}